# Comment Diffusons-nous sur les Réseaux Sociaux ?

Erick STATTNER

*Maitre de Conférences en Informatique*

Université des Antilles

*erick.stattner@univ-ag.fr*

*Information, influences et propagation via les médias en ligne*

**Communication de recherche**

**Mots-clés:** Réseaux complexes; Média sociaux; Diffusion; Analyse et fouille de données; Comportement humain

## Resume

L'émergence des nouveaux moyens de communication tels que les blogs, les journaux en ligne ou les réseaux sociaux nous permettent d'aller plus loin dans la compréhension des comportements humains. En effet, ces espaces d'échange publiques sont aujourd'hui fermement ancrés dans nos sociétés modernes et s'avèrent être de puissants capteurs des comportements sociaux et des mouvements d'opinions. Dans ce travail, nous nous intéressons à la diffusion d'informations et cherchons à comprendre quelles sont les conditions dans lesquelles une personne décide de s'exprimer sur un sujet. Nous proposons pour cela un ensemble de mesures qui visent à caractériser les comportements de diffusion. Nos mesures ont été utilisées sur les messages liés à deux évènements ayant eu lieu au mois de Janvier 2015: la présentation par Microsoft d'un nouveau casque de réalité virtuelle et l'élection d'un parti politique de gauche radical en Grèce.

# I. INTRODUCTION

La capacité qu'ont aujourd'hui les medias sociaux à agréger les écrits, les opinions et les pensées de l'humanité en font des outils puissants pour comprendre l'inconscient collectif et identifier les tendances émergentes. De nombreux travaux ont par exemple montré qu'un réseau social tel que Twitter est un extraordinaire capteur de comportements sociaux qui renferme en son sein des indicateurs qui, une fois traités et fouillés, permettent d'aborder des questions très larges telles que l'incidence d'une épidémie dans une population (Gomide et al., 2011), le résultat d'une élection politique (Tumasjan et al., 2010), l'évolution des indices de marché boursier (Bollen et al., 2011) ou la prévision du revenu d'un film au box office (Asur et Huberman, 2010).

Dans ce travail, nous nous intéressons au problème de la diffusion d'informations et cherchons à comprendre quelles sont les conditions dans lesquelles une personne décide de s'exprimer sur un sujet. Ce problème est particulièrement intéressant puisque l'émergence de ces mêmes médias sociaux a également redéfini nos comportements de communication. D'une transmission d'informations de personne à personne, longtemps qualifiée de "*bouche à oreille*", nous sommes aujourd'hui dans l'ère de la communication impersonnelle, où nous retransmettons de l'information de manière asynchrone sans contact personnalisé avec les potentiels destinataires.

Si de nombreux modèles de diffusion ont été proposés dans la littérature, tels que modèles à compartiments (Mollison, 1995; Diekmann and Heesterbeek, 2000), des modèles à métapopulation (Grenfell et al., 1997, Colizza et al. 2007) ou des modèles basés sur les réseaux (Salathe and Jones, 2010; Stattner et al., 2012), ils visent à reproduire les tendances globales observées, à travers des règles de transmission simplifiées qui ne reflètent pas toute la complexité des comportements individuels véritablement impliqués dans ces processus. Par exemple, en étudiant la diffusion de fichiers dans un réseau pair-à-pair, Albano et al. (Albano et al., 2010) ont montré que le modèle de diffusion classique SI (*Susceptible-Infecté*) ne reproduit pas fidèlement le phénomène.

Le réseau social Twitter est un bon cas d'étude pour aborder le problème de la diffusion puisque tous les sujets de société y sont abordés. Il possède une grande variété d'utilisateurs: professionnels, particuliers, politiciens, associations, syndicats, entreprises, etc. La plateforme est, de plus, utilisée par 200 millions d'utilisateurs et 500 millions de messages y sont publiés chaque jour[1]. Ainsi, les tendances observées sur ce réseau fournissent des indicateurs pertinents pour comprendre comment l'information se transmet et se propage dans une communauté.

Dans ce travail, nous utilisons donc Twitter comme média support et cherchons à comprendre les comportements en termes de diffusion. Notre objectif est d'apporter des éléments de réponse à des questions telles que:
**(Qi)** Un individu diffuse t-il parce qu'il se sait très écouté?
**(Qii)** Un individu diffuse t-il parce qu'il est inondé de messages par sa communauté?
**(Qiii)** Combien de temps attend un individu avant de s'exprimer?

---

[1] source: www.blogdumoderateur.com/chiffres-twitter/

Pour répondre à ces questions, nous proposons un ensemble de mesures qui visent à décrire le processus globalement et localement. Nous commençons par utiliser l'API Twitter (Yamamoto, 2010) pour collecter tous les messages émis sur un sujet donné et pour obtenir l'état de voisinage d'un individu à chaque fois qu'il s'exprime sur le sujet ciblé. Nos mesures sont parallèlement calculées sur le jeu de données chaque fois qu'un nouveau message est identifié.

Notre approche a été utilisée pour étudier pendant trois jours, les messages liés à deux événements ayant eu lieu en Janvier 2015. Le vendredi 23, la présentation par Microsoft d'un nouveau casque de réalité virtuelle appelé *HoloLens* et le Samedi 25, l'élection du parti politique de gauche radical *Syriza* en Grèce. Ces événements ont été choisis pour leur nature différente. D'une part un sujet technologique qu'on suppose être relativement neutre, et de l'autre un sujet politique, naturellement plus clivant, et susceptible de générer des discussions actives au sein du réseau. Les résultats obtenus montrent qu'il existe une forte hétérogénéité dans les comportements individuels de diffusion. Enfin, notre approche a été implémentée dans un outil graphique, appelé ThotS, qui permet de mener une analyse en temps réel sur toute sorte de sujets ciblés sur Twitter.

Ce papier est organisé en sept sections. La Section 2 présente les travaux pionnés menés sur l'analyse de données issues de Twitter pour aborder différents phénomènes. La Section 3 est consacrée à la méthodologie que nous proposons pour caractériser le processus de diffusion. Les Sections 4 et 5 présentent les résultats obtenus sur deux cas d'étude. La Section 6 décrit l'outil graphique ThotS qui implémente l'approche proposée. Enfin, nous concluons dans la Section 7 et présentons nos travaux futurs.

## II. DE TWITTER A LA CONNAISSANCE

Twitter offre une gigantesque base de données, qui couvre à la fois les messages publiés sur différentes catégories de sujets et des informations personnelles sur les utilisateurs du système. Bien que les travaux menés sur l'analyse des données issues de Twitter soient relativement récents, ils ont déjà permis d'aborder de nombreux phénomènes du monde réel.

Par exemple, Guille et Favre (Guille et Favre, 2014) ont récemment proposé une méthode qui analyse en temps réel les messages postés sur Twitter afin de détecter l'apparition de pics de popularité sur des sujets. Contrairement à l'approche classique appelée *Peaky Topics* (Shamma et al., 2011), qui repose sur une augmentation soudaine de la fréquence d'apparition d'un mot ou d'une expression, la méthode proposée est basée sur la notion de liens dynamiques, c'est-à-dire des messages faisant référence à d'autres utilisateurs. Les auteurs montrent ainsi que la fréquence des références est un indicateur du niveau d'intérêt que génère le sujet.

Dans un autre contexte, Sakaki et al. (Sakaki et al., 2010) s'intéressent à la détection d'événements ciblés et proposent un algorithme capable de surveiller les messages postés sur Twitter et de détecter ces événements. Ils se concentrent en particulier sur la détection des tremblements de terre et ils montrent que leur système détecte les tremblements de terre rapidement avec une forte probabilité (96% des tremblements de terre recensés par l'Agence météorologique japonaise (JMA)). Leur approche a été mise en œuvre dans un système d'alertes qui fournit des notifications beaucoup plus rapidement que les annonces envoyées par la JMA.

Sur le même principe, nous pouvons citer les travaux d'analyse des messages postés sur Twitter menés par Gomide et al. (Gomide et al., 2011) pour évaluer l'incidence de l'épidémie de dengue au Brésil en termes de perception du public et d'évolution spatio-temporelles, ou les travaux de Velardi et al. (Velardi et al., 2014) pour la détection précoce et l'analyse d'une épidémie dans une population.

Plus généralement, les messages issus de Twitter ont été utilisés pour étudier différents types d'événements du monde réel. Tumasjan et al. s'intéressent par exemple à l'élection fédérale allemande (Tumasjan et al., 2010). Ils observent que les opinions présentes dans les messages mentionnant un parti reflètent le résultat des élections. Dans (Bollen et al., 2011), les auteurs observent que l'humeur publique extraite de messages sur Twitter peut améliorer la prédiction des indices de marché boursier. Asur et Huberman vont plus loin et déclarent que les médias sociaux tels que Twitter peuvent "*prédire le futur*" (Asur et Huberman, 2010). Ils démontrent leur assertion en analysant les messages de Twitter pour prédire les recettes de films au box-office.

Sakaki et al. tentent de généraliser ce type d'approche en soutenant que des médias sociaux tels que Twitter peuvent être exploités pour étudier tout phénomène présentant trois propriétés principales (Sakaki et al., 2011): (i) une apparition à grande échelle, c'est-à-dire qu'un très grand nombre d'utilisateurs sont concernés, (ii) une influence sur la vie quotidienne des gens, qui est la raison pour laquelle ils décident de poster un ou plusieurs messages sur le sujet, et (iii) une dimension spatiale et temporelle, de sorte qu'une analyse spatio-temporelle soit possible. Ces propriétés sont couvertes par de nombreux phénomènes naturels tels que des tremblements de terre, des ouragans, des tsunamis, etc., mais aussi des événements sociaux tels que des événements sportifs, des conférences, des élections, etc.

## III. MESURER LA DIFFUSION

Le site de micro-blogging Twitter (Le microblog est un dérivé du blog, qui permet de publier des articles courts) peut être vu comme un immense réseau orienté (c'est-à-dire qu'un lien de A vers B, n'implique pas nécessaire l'existence d'un lien de B vers A) dans lequel chaque utilisateur (nœud du réseau) a un ensemble d'abonnés qui le "*suivent*" appelés *Followers* et un ensemble d'utilisateurs auxquels il est abonné et qu'il "*suit*", appelés *Followings*. L'ensemble des followers et des followings constitue les liens du réseau. La notion de "*suivi*" fait ici référence à l'intérêt pour le contenu publié par un utilisateur.

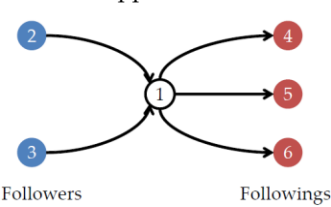

**Figure 1. Exemple de followers et followings de l'utilisateur 1**

La Figure 1 montre un exemple du réseau induit par Twitter à travers les liens de suivi. Les utilisateurs 2 et 3 sont les Followers de l'utilisateur 1 et les utilisateurs 4, 5 et 6 sont ses Followings.

Chaque utilisateur peut publier des messages, appelés *Tweets*, qui correspondent à des textes courts qui n'excedent pas 140 caractères. Ces messages sont généralement des nouvelles, des opinions sur un sujet donné ou des informations personnelles sur la vie de l'utilisateur. Ils peuvent également contenir des liens vers des images, des vidéos ou des articles. Les tweets sont affichés sur la page de profil de l'utilisateur et sont présentés à ses followers. Un ensemble de meta-données est également associé à chaque tweet telles que la date, le lieu, l'auteur, les liens et les hashtags utilisés, etc.

Twitter offre également la possibilité de retransmettre un tweet publié par un utilisateur, ce concept est appelé le *retweet*. Le retweet est un moyen populaire de propager un contenu intéressant à ses Followers. Ainsi, chaque fois qu'un utilisateur publie un message, ses followers sont susceptibles de le lire et de le retransmettre à l'attention de leurs propres followers, soit en retweetant le message original, soit en publiant un nouveau tweet sur le même sujet.

Notre objectif est de comprendre quelles sont les conditions dans lesquelles une personne décide de s'exprimer sur un sujet. Nous proposons pour cela un ensemble de mesures que nous utilisons pour répondre aux questions (**Qi**), (**Qii**) et (**Qiii**) présentées dans l'introduction. Plus concrètement, nous utilisons l'API Twitter pour capturer en temps réel l'ensemble des messages émis sur un sujet donné. Chaque fois qu'un message est publié sur le sujet ciblé, nous maintenons à jour deux types d'indicateurs :

1. **Indicateurs globaux**, qui visent à décrire globalement le processus et son évolution dans le temps :
    - *NbTw, NbRTw, NbUs*: correspondent respectivement aux nombres de tweets, de retweets et d'utilisateurs enregistrés sur la période d'étude.
    - *AVG(Tw/Us), AVG(RTw/Us)* : correspondent respectivement aux nombres moyens de tweets et de retweets publiés par utilisateur.
    - *AVG($T_{tw}$), AVG($T_{rtw}$):* correspondent respectivement à la durée moyenne (en secondes) entre deux tweets et deux retweets publiés sur le sujet ciblé.
2. **Indicateurs locaux**, qui visent à caractériser l'environnement d'un utilisateur *u* donné la première fois qu'il publie un message sur le sujet ciblé.
    - *$NbFe_u$*: représente le nombre de followers de l'utilisateur *u*.
    - *$NbFgP_u$*: correspond au nombre de followings de l'utilisateur *u* qui ont publié un message avant lui sur le sujet ciblé.
    - *$NbT_u$*: est le nombre total de tweets publiés sur le sujet par l'utilisateur *u*.
    - *$TotalR_u$*: correspond au nombre total de messages envoyés par les followings de l'utilisateur *u*.
    - *$Elapsed_u$*: est le temps écoulé avant que l'utilisateur *u* s'exprime pour la première fois (en heures).

Ces mesures ont été utilisées pour étudier deux événements ayant eu lieu en Janvier 2015: (a) le vendredi 23, la présentation par Microsoft d'un nouveau casque de réalité virtuelle appelé *HoloLens* et (b) le Samedi 25, l'élection en Grèce du parti politique de gauche radical *Syriza*. Les messages sur ces deux événements ont été collectés sur 72 heures. Nous précisons également que les résultats sont présentés avec le fuseau horaire de Paris en France (UTC+1).

Les messages sur l'événement HoloLens ont été obtenus en capturant tous les tweets contenant les mots-clés "HoloLens" ou "Holo Lens". De même, l'ensemble des messages sur Syriza a été obtenu en capturant tous les messages contenant les mots-clés "Syriza" ou "Tsipras". Enfin, nous précisons que la méthode de collecte n'est pas sensible à la casse.

## IV. COMPORTEMENTS MACROSCOPIQUES

Dans une première approche, nous nous sommes concentrés sur le comportement global des phénomènes étudiés. Notre objectif était de comprendre comment évolue avec le temps la diffusion des messages, au travers des indicateurs globaux proposés dans la section précédente. La Figure 2 montre, pour chacun des évènements, l'évolution de ces indicateurs en fonction du temps : (a) nombre de tweets, de retweets et de nouveaux utilisateurs (*NbTw*, *NbRTw* et *NbUs*), (b) nombre moyen de tweets et de retweets par utilisateur ($AVG(Tw/Us)$ et $AVG(RTw/Us)$) et (c) temps moyen (en secondes) entre deux tweets et retweets ($AVG(T_{tw})$ et $AVG(T_{rtw})$).

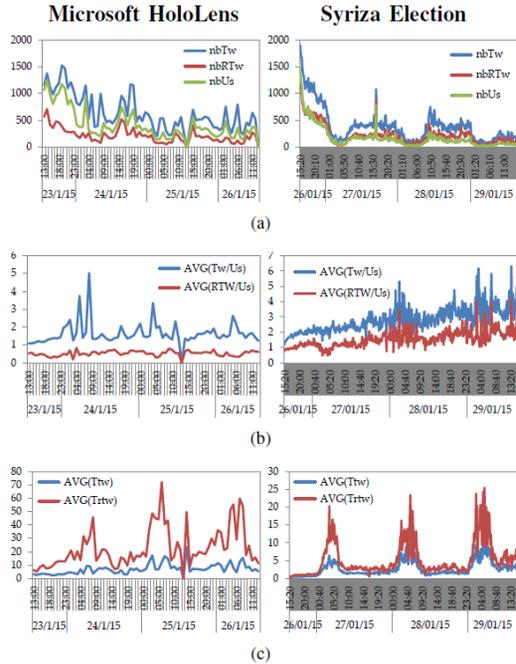

Nous observons tout d'abord que les deux évènements n'ont pas la même amplitude: 46 317 tweets, 1 5633 retweets et 30 629 utilisateurs ont été collectés pour HoloLens alors que pour Syriza nous dénombrons 168 255 tweets, 94 780 retweets et 74 101 utilisateurs.

Nous pouvons également observer que dans les deux cas, l'intérêt pour le sujet diminue avec le temps (cf. Figure 2 (a)). Cela se traduit par une baisse au cours du temps du nombre de messages (tweets et retweets) et du nombre de nouveaux utilisateurs s'exprimant. Il est également intéressant de noter la différence dans le comportement des utilisateurs pour ces deux évènements. En effet, pour l'événement HoloLens, le nombre de nouveaux utilisateurs est plus élevé que le nombre de retweets. Ce résultat laisse à penser que pour le sujet technologique, les utilisateurs ont peu utilisé la fonctionnalité de retweet et ont préféré soumettre leur propre message.

**Figure 2. Evolution des indicateurs globaux**

En ce qui concerne le nombre moyen de messages (tweets et reteweets) par utilisateur (cf. Figure 2 (b)), nous observons que pour les deux évènements, le nombre de tweets par utilisateur est toujours supérieur à celui des retweets. Ce résultat traduit le fait que les utilisateurs tendent à commenter eux-mêmes le sujet plutôt que retweeter un message déjà soumis. En revanche, bien que le nombre d'utilisateurs décroisse avec le temps (observé sur la Figure 2 (a)), nous notons que sur le sujet politique, ceux qui poursuivent les discussions sont beaucoup plus actifs. Cela se traduit en particulier par une hausse du nombre moyen de messages par utilisateur, qui n'est pas observée dans le cas du sujet technologique. Ainsi, alors que la communauté autour du sujet politique décroit, les utilisateurs qui maintiennent les discussions sont, eux, plus actifs. Cela peut s'expliquer par la nature politique du sujet, qui est plus enclin à générer des échanges actifs du type *"argument/contre argument"*.

Les résultats obtenus concernant le temps moyen entre deux messages sont très intéressants (cf. Figure 2 (c)). Dans les deux événements, nous notons que le temps écoulé entre deux retweets est toujours plus élevé que celui observé entre deux tweets. Cela confirme notre précédente observation selon laquelle le nombre de retweets est moins important. De plus, sur le cas Syriza, des motifs répétitifs peuvent être observés, durant lesquels le temps entre deux messages est très faible, puis croît considérablement avant de rechuter. Les périodes où le temps est le plus long correspondent aux nuits en Europe. On peut supposer que les citoyens européens, qui sont les plus concernés par l'événement, et donc les plus présents dans les discussions, sont moins actifs durant ces périodes. Ces motifs ne s'observent pas dans le cas de HoloLens. Cela peut être expliqué par le fait que le sujet de la réalité augmentée est un sujet qui va au-delà des zones géographiques et qui fait réagir des communautés présentes dans le monde entier.

# V. VERS UNE DIFFUSION INDUITE PAR DES COMPORTEMENTS INDIVIDUELS

Dans une deuxième approche, nous abordons le problème d'un point de vue local, en étudiant les situations dans lesquelles un individu est le plus susceptible de s'exprimer sur un sujet. Nous cherchons, en particulier, à répondre aux questions soulevées dans l'introduction: **(Qi)** Un individu diffuse t-il parce qu'il se sait très écouté? **(Qii)** Un individu diffuse t-il parce qu'il est inondé de messages par sa communauté? **(Qiii)** Combien de temps attend un individu avant de s'exprimer?

Pour répondre à ces questions, nous nous intéressons à la distribution des indicateurs locaux présentés à la Section III. La figure 3 montre les résultats obtenus pour les deux évènements : (a) nombre de tweets envoyés ($NbT_u$), (b) nombre de followers ($NbFe_u$), (c) nombre de followings qui se sont exprimés avant ($NbFgP_u$), (d) nombre total de messages reçus avant de publier son premier message ($TotalR_u$) et (e) le temps écoulé avant tweeting ($Elapsed_u$).

Nous pouvons tout d'abord observer que la majorité des utilisateurs n'a publié qu'un seul message (cf. Figure 3 (a)). En effet, environ 80% des utilisateurs n'ont envoyé qu'un message dans le cas de HoloLens, contre 65% dans le cas de Syriza. Ce résultat confirme une observation déjà faite dans la section précédente, selon laquelle la majorité des utilisateurs publie finalement peu de messages.

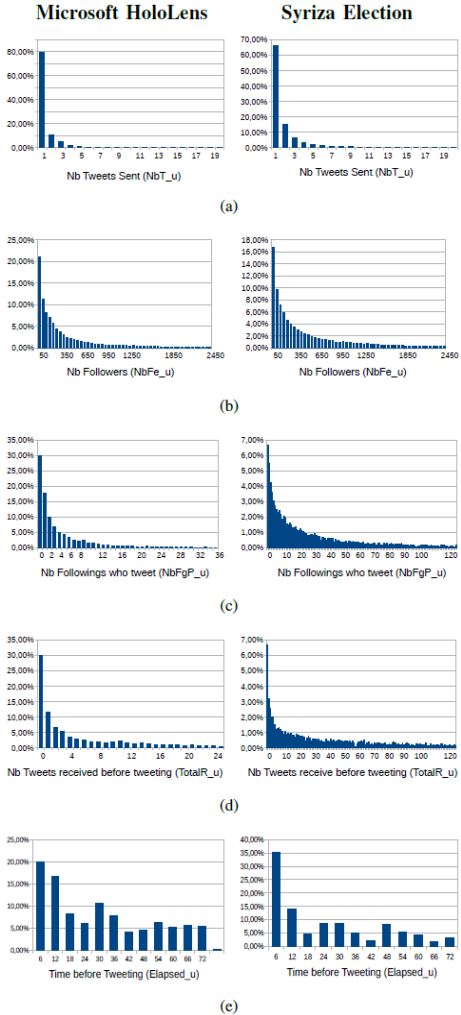

Figure 3. Distribution des indicateurs locaux

En ce qui concerne les résultats obtenus pour le nombre de followers des individus qui publient (cf Figure 3 (b)), les tendances sont similaires pour les deux évènements. En effet, la distribution suit une loi de puissance ce qui traduit le fait que la grande majorité des personnes qui se sont exprimées ont peu de followers. Il est, de plus, particulièrement intéressant de noter qu'un pourcentage élevé d'utilisateurs ont posté un message alors qu'ils n'ont aucun follower. Par exemple, dans le cas d'HoloLens, environ 20% des individus ont publié un message alors qu'il ne sont suivis par aucun follower (16% dans le cas de Syriza).

**Ainsi, ces résultats n'accréditent pas la thèse de l'influence du nombre de followers sur le comportement de diffusion comme formulée par la question (Qi).**

En revanche, les résultats obtenus pour le nombre de followings à avoir publié un message avant que l'utilisateur ne poste lui-même un message sont assez différents pour les deux évènements (cf. Figure 3 (c)). En effet, pour le cas HoloLens, environ 30% des utilisateurs ont publié un message alors qu'aucun de leur followings n'en avait soumis (environ 7% pour Syriza). Ce résultat suggère que, dans le cas du sujet politique étudié, les individus ont tendance à attendre les réactions des personnes qu'ils suivent avant de réagir. Les mêmes tendances sont observées si on s'intéresse au nombre total de messages envoyés par les followings d'un utilisateur avant qu'il publie lui-même un message (cf. Figure 3 (d)). En effet, dans le cas d'HoloLens 30% des utilisateurs ont publié un message sans qu'aucun de leur following n'en ait posté (6,5% for Syriza).

**Ainsi, les résultats obtenus ne montrent pas clairement que les personnes attendent d'être submergées de messages par leur communauté avant de réagir, comme suggéré par la question (Qii). Nous observons en effet une forte différence dans les comportements de diffusion selon les sujets étudiés. Si pour le sujet technologique les réactions ne semblent pas être dépendantes des messages reçus,**

**pour le sujet politique, les utilisateurs semblent attendre les premières réactions de leur communauté avant de réagir.**

Enfin, en ce qui concerne le temps écoulé avant que les utilisateurs postent leur premier message (cf. Figure 3 (e)), nous observons que pour les deux évènements, la majeure partie des premières réactions surviennent durant les premières 24 heures. Elles décroissent également avec le temps. Par exemple, pour le cas HoloLens, la moitié des messages a été postée au cours des 24 premières heures (environ 62% pour Syriza). Évidemment, le nombre de premières réactions diminue avec le temps, ce qui reflète une perte d'intérêt pour le sujet déjà observée dans nos résultats précédents. Il est d'ailleurs intéressant de noter que certains utilisateurs postent des messages pour la premiere fois jusqu'à 72 heures après le début de l'évènement : environ 5% des individus pour HoloLens, contre 2% pour Syriza.
**Ces résultats apportent ainsi des éléments de réponse à la question (Qiii), puisque nous observons que les premières réactions ont lieu dans les premières heures après le début de l'événement et diminuent fortement avec le temps.**

Pour aller plus loin dans la compréhension de ces processus, nous avons cherché à vérifier s'il existe des corrélations dans certains comportements de diffusion. Plus particulièrement, nous cherchions à savoir si ceux qui postent le plus de messages sont ceux qui ont le plus de followers, ou sont ceux qui ont été exposés au plus grand nombre de messages de leur followings. La figure 4 montre les résultats obtenus entre: (a) le nombre total de messages postés ($NbT_u$) et le nombre de followers ($NbFe_u$) ; (b) le nombre de messages postés ($NbT_u$) et le total des messages publiés par les followings ($TotalR_u$).

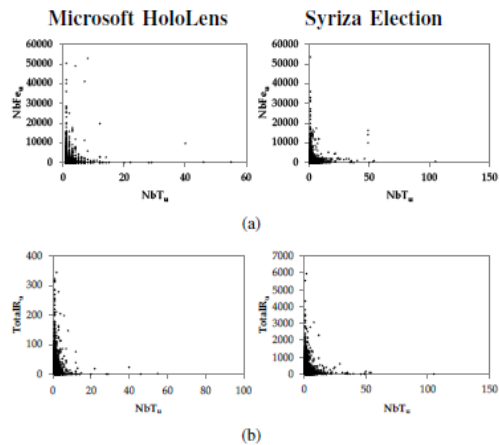

Figure 4. Correlations d'attributs individuels

Nous pouvons tout d'abord observer que les individus qui postent peu de messages (moins de 20) sont très hétérogènes en termes de followers (cf. Figure 4 (a)). Certains ont très peu de followers, alors que d'autres en ont énormément. Cependant, si nous nous concentrons sur les individus qui se sont exprimés le plus, la population est beaucoup plus homogène. En effet, les individus les plus actifs ne possèdent finalement que très peu de followers.
**Ce résultat confirme, pour les deux évènements étudiés, la non-influence du nombre de followers dans la quantité de messages envoyés par un utilisateur comme le suggère la question (Qi).**

Des tendances similaires sont observables en ce qui concerne les corrélations entre le nombre de messages postés et le total des messages publiés par les followings (cf. Figure 4 (b)). En effet, les individus qui s'expriment le moins sont très hétérogènes, puisque nous y trouvons des individus dont les followings peuvent avoir envoyé très peu ou énormément de messages. Cependant, la population composée des individus les plus actifs est plus homogène, dans le sens où on y retrouve uniquement des individus dont les followings ont publié peu de messages.

## VI. THOTS: UN OUTIL DE COLLECTE ET D'ANALYSE

La méthodologie proposée a été implémentée dans un outil graphique appelé *ThotS* [2] (cf. capture d'écran Figure 5). ThotS a été conçu pour collecter et analyser en temps réel les messages publiés sur Twitter. L'outil met en œuvre les mesures globales et locales proposées pour décrire le phénomène et travaille sur trois axes.

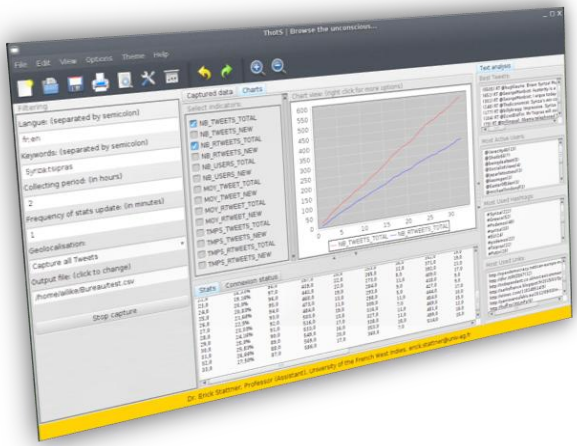

**Figure 5. Capture d'écran de ThotS**

**1. Collecte**. ThotS capture en temps réel tous les messages publiés sur un sujet donné. Les messages peuvent être filtrés selon différents critères : mot-clés, langue, position, etc. (cf. panneau gauche Figure 5). Tous les messages collectés sont sauvegardés dans un fichier texte pour une analyse future ou dans le but d'une analyse avec des logiciels d'analyse spécialisés dans la visualisation, l'extraction de connaissances, le big data, la modélisation, etc.

**2. Exploration**. ThotS explore les données collectées selon les indicateurs globaux et locaux présentés à la Section III. Les indicateurs sont calculés et accessibles en temps réel, à travers des tableaux récapitulatifs (cf. panneau bas Figure 5) et des graphiques montrant leur évolution au cours du temps (cf. panneau central Figure 5). Les graphiques sont générés en utilisant l'API développée en JAVA JFreeChart (Gilbert, 2002). L'outil permet également de choisir les indicateurs à afficher et de les superposer sur un même graphique.

**3. Extraction de connaissances.** ThotS analyse enfin jusqu'au contenu des messages pour en extraire toute l'essence, en fournissant une connaissance utile et directement opérationnelle: meilleurs tweets, mots les plus utilisés, utilisateurs les plus actifs, liens les plus cités, etc. (cf. panneau droite Figure 5).

---

[2] ThotS: http://erickstattner.com/thots-analytics/

## VII. CONCLUSION ET PERSPECTIVES

Dans cet article, nous avons abordé le problème de la diffusion d'informations sur les medias sociaux. Contrairement aux approches traditionnelles qui se concentrent sur différents types de modèles de diffusion pour comprendre ces phénomènes, nous avons abordé ici le problème d'un autre point de vue en étudiant la diffusion à partir de situations réelles survenant sur le site de micro-blogging Twitter.

La première étape de notre travail a consisté à proposer un ensemble de mesures qui visent à décrire la diffusion globalement et localement. Ces mesures ont ainsi pu mettre en lumière certaines dynamiques individuelles impliquées dans la diffusion de messages liée à deux évènements ayant eu lieu en janvier 2015.

Cependant, bien que ces résultats aient permis d'apporter des éléments de réponse sur les comportements de diffusion, nous devons garder à l'esprit qu'ils ne peuvent être généralisés, puisque nous avons également observé que la sémantique du sujet ciblé est une dimension à prendre en compte pour comprendre pleinement ces phénomènes. Ainsi, à court terme, nous prévoyons d'étendre notre étude à d'autres types de sujets, dans l'objectif de comprendre l'impact de la sémantique du sujet sur les comportements.

Les résultats obtenus soulèvent également de nombreuses questions intéressantes concernant la modélisation. En effet, la plupart des modèles de diffusion supposent que la diffusion est induite par des règles très simples. A long terme, nous voulons exploiter les tendances observées pour proposer des modèles plus réalistes tenant compte des comportements individuels et la sémantique du phénomène.

Enfin, ce travail ouvre également des perspectives intéressantes en termes de modélisation prédictive. En effet, les données recueillies pourraient être utilisées pour construire des modèles prédictifs qui pourraient être en mesure de prédire, pour un utilisateur donné, s'il transmettra une information.

## BIBLIOGRAPHIE